\documentclass{an}
\usepackage{amsmath}
\usepackage{amssymb}
\usepackage{graphicx}
\usepackage{times}
\usepackage{epstopdf}
\usepackage{natbib}

\begin{document}

\Pagespan{1}{4}
\Yearpublication{2012}
\Yearsubmission{2012}
\Month{Month of publication}
\Volume{Volume}
\Issue{Issue}
\DOI{DOI}
\title{Photo-Met: a non-parametric method for estimating stellar metallicity from photometric observations}
\author{Gy\"ongyi Kerekes\inst{1}, Istv\'an Csabai\inst{1}, L\'aszl\'o Dobos\inst{1}, M\'arton Trencs\'eni\inst{1}}
\institute{Dept.\ of Physics of Complex Systems, E\"otv\"os Lor\'and University, P\'azm\'any P.\ s\'et\'any 1/A, Budapest, 1117, Hungary}
\received{15/11/2012} \accepted{A-date}
\publonline{later}
\abstract{Text of abstract}

\authorrunning{Kerekes, Csabai, Dobos et al.}
\titlerunning{Photometric metallicity estimation method}


\abstract{Getting spectra at good signal-to-noise ratios takes orders of magnitudes more time than photometric observations. Building on the technique developed for photometric redshift estimation of galaxies, we develop and demonstrate a non-parametric photometric method for estimating the chemical composition of galactic stars. We investigate the efficiency of our method using spectroscopically determined stellar metallicities from SDSS DR7. The technique is generic in the sense that it is not restricted to certain stellar types or stellar parameter ranges and makes it possible to obtain metallicities and error estimates for a much larger sample than spectroscopic surveys would allow. We find that our method performs well, especially for brighter stars and higher metallicities and, in contrast to many other techniques, we are able to reliably estimate the error of the predicted metallicities.}

\maketitle

\section{Introduction} \label{sec:intro}

Since their introduction \citep{Baum1962, Connolly1995} photometric redshift estimation (photo-z) methods have been developed into a widely accepted and used technique for galaxies. This is not surprising since most modern optical surveys like the Sloan Digital Sky Survey (SDSS), Pan-STARRS or the planned Large Synoptic Telescope (LSST) are optimized for getting wide-band photometry for large number of objects and the more time consuming spectroscopy is done only for a fraction of the sources, if at all. The main idea behind photo-z is based on the fact that spectral information (redshift) is conserved at some level when the high resolution, information rich spectra are projected to the low-dimensional wide-band photometric space. Following the footsteps of photo-z, in this work we introduce {\it photo-Met}, a non-parametric photometric metallicity estimation method. To demonstrate and evaluate its capabilities, we apply photo-Met to 12,000 stars of SDSS. We perform the accuracy analysis for SDSS stars in various magnitude intervals where both spectral and photometric observations are available and conclude that the major limitation of the method is the accuracy of photometry and the lack of a good training set for certain metallicity ranges. Compared to many other estimation methods an advantage of our technique is that beyond metallicity estimates, it yields reliable uncertainties as well. The method needs no input parameters, there are no cuts in the stellar type and it works relatively fast.

\section{The metallicity estimation algorithm} \label{sec:method}

Accurate metallicity, surface gravity, effective temperature, radial velocity and other parameters of individual stars are usually determined from spectroscopic observations or sometimes from narrow band photometry. Although in a rather condensed way, most of this information is contained by broad-band photometric data. Since photometric data are available for far more stars than spectroscopic observations, the idea to estimate the main stellar parameters using only broad-band filters has already been investigated in numerous studies. Recently, \cite{Ivezic2008} used third-order polynomials to fit $u-g$ and $g-r$ SDSS colors of start in order to determine metal abundances of 200,000 F and G type stars. Note, that a similar polynomial fitting technique was one of the earliest successful methods of photo-z \citep{Connolly1995} as well.

The approach used in this study is a non-parametric, so called {\it empirical} technique. We use broad-band photometry as input and, based on the information gathered from a {\it training set} of stars (for which both spectroscopic metallicity and photometric colors are known), we create an empirical relation between broad-band colors and metallicity. This empirical relation, which is more like a computer algorithm than an analytical function, is then used to estimate unknown metallicities for the {\it query set} of stars. The method relies on the findings of \cite{Fukugita2011}, namely that, for most of the stellar loci, different metallicities result in different colors. Hence, knowing the colors of a star, one can readily give an estimate on its metal abundance. Our training set is based on large number of stars (∼230,000) with abundances calculated from medium-resolution SDSS spectra \citep{Beers2006, Allende2006, Lee2008a}. 

\subsection{K-nearest neighbour kd-tree method} \label{sec:kd}

Based on the observational result cited above, our method assumes that objects which are located close in the color-color space have very similar metal abundances. For every query object our algorithm searches the training set and determines $k$ training set objects that are the nearest to the query object in the 4 dimensional ($u-g$, $g-r$, $r-i$, $i-z$) color index space of SDSS. Then multi-linear regression is used to get an estimate on metallicity, in the following form.
\begin{equation}
\left[ \frac{Fe}{H} \right](\vec{x}) = c_0 + \vec{c_1}\,\vec{x}
\end{equation}
where $c_0$ is a constant, $\vec{c_1}$ is a vector of coefficients and $\vec{x}=[u-g, g-r, r-i, i-z]$ is a 4-D vector formed from the color indices. K-nearest neighbour search in large point sets is computationally challenging. We use multi-dimensional kd-tree indexing to speed up neighbour lookup \citep{Csabai2007}. The code is implemented in C/C++ originally for photometric redshift estimation and required small modification only to be applicable for this problem.

Due to the fat tail error distribution of both photometric colors and metallicity values, outliers are frequent, namely, stars with different metallicities will occupy the same loci in the color index space. To resolve this degeneracy and also to make the method robust, we discard neighbours with metallicities more than $3\sigma$ away from the fitted hyperplane described by $c_0$ and $\vec{c_1}$. The remaining points are used to fit the hyperplane repeatedly until convergence. We use $k=50$ nearest neighbors, the optimal value depends on the size of the training set and its photometric error properties and has moderate effect on the accuracy of the estimation. 
%Based on the final regression we calculate the metal abundance estimation and we are able to calculate the error of the estimation, too.

\subsection{Training Set}

SDSS was designed to survey the northern sky and originally concentrated mostly on extragalactic sources. It turned out, however, that studying galactic stars is also very productive with the same instrument. The Sloan Extension for Galactic Understanding and Exploration (SEGUE) was done during SDSS~II and was aimed to obtain spectra for 240.000 stars in the Galaxy with a typical chemical abundance accuracy of 0.3 dex and photometric errors of $~1\%$ in the g, r, i, and z bands, and a slightly worse of $~2\%$ in the u band. These medium resolution spectra are of sufficient quality to determine stellar parameters including metallicity, surface gravity, and effective temperature. Stellar parameters are determined using different methods implemented in an automated processing pipeline called the SEGUE Stellar Parameter Pipeline (SSPP, \cite{Beers2006}). A detailed discussion of these methods and their performance can be found in \cite{Allende2006, Lee2008a, Lee2008b}, and in \cite{Yanny2009}.

All stellar spectral types are represented as the survey was aimed to sample stars the stellar populations of the spheroid substructure, the thin and the thick disk. A small subset of the spectroscopic fibres was devoted to unusual stars, including those thought to have low metallicities of ($[M/H] < -2$), or being otherwise unusual based on their colors and peculiar velocities. In addition, star clusters with a variety of ages and metallicities were also targeted.

To construct our training set we used the extinction corrected PSF magnitudes and took the spectroscopically  determined $[Fe/H]$ values from the SSPP. We applied color cuts to exclude non-stellar objects, a limit on maximum extinction and a cut in the $g$ magnitude:

\begin{itemize}
\item[--]
interstellar extinction in r band $< 0.3$
\item[--]
$14 < g < 20$ 
\item[--]
$0 < u-g < 3.5$ and
$-0.6 < g-r <2$ and \\
$-1 < r-i < 2$
\item[--] exclusion of quasars (Finlator 2000) with colors of: \\
$ u-g < 0.4$ and
$-0.1 < g-r < $ and 
$r-i< 0.5$
\end{itemize}

As a result, we got 233,448 stars with 4 color indices and a $[Fe/H]$ parameter and its error which altogether served as the training set.

\begin{figure}
\includegraphics[scale=0.55]{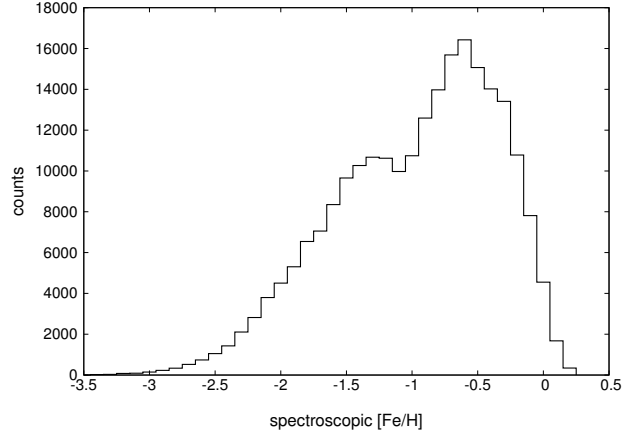}
\caption{The distribution of metallicity in the training set. The low metal abundance stars are somewhat under-represented, for a better training set a flat distribution would be ideal.}
\label{fig:TS}
\end{figure}

\begin{figure*}
\includegraphics[scale=1.35]{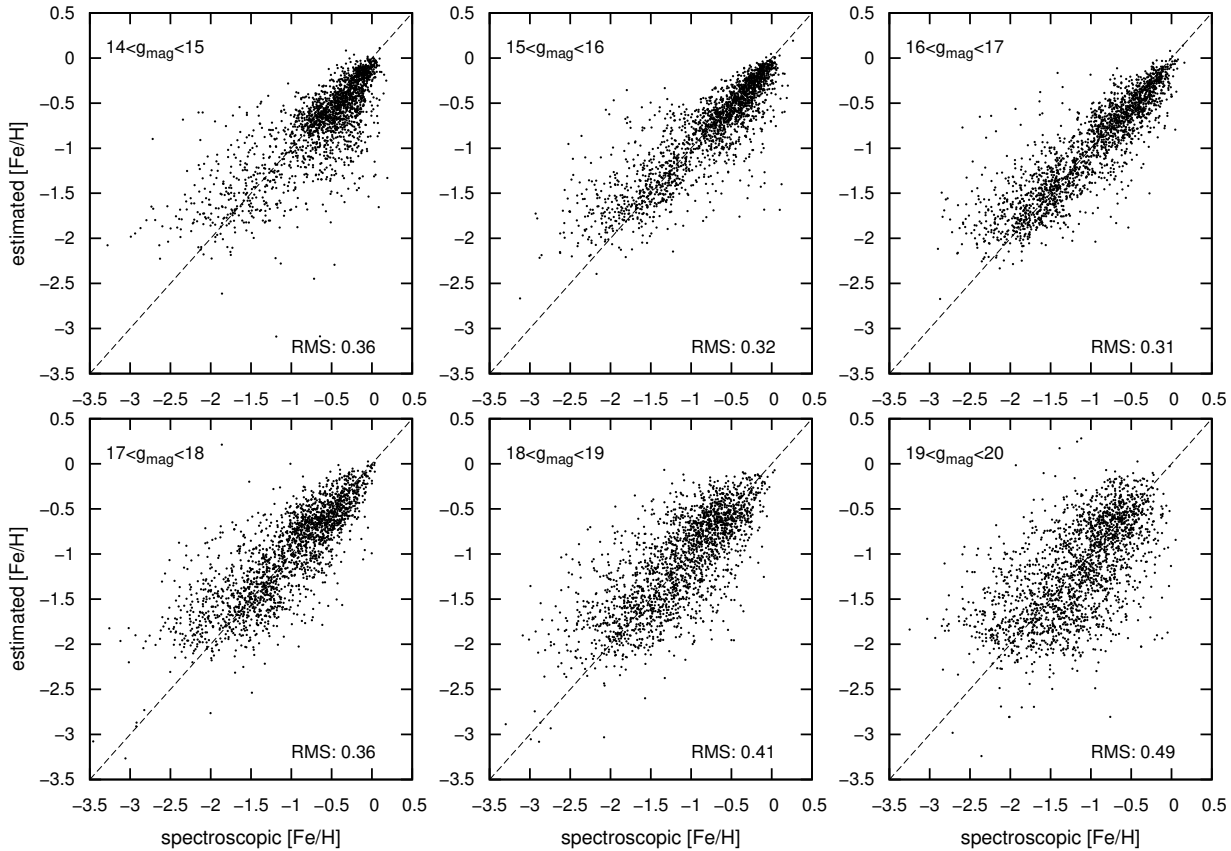}
\caption{Photometric metallicity estimation of spectroscopically measured stars from SDSS DR7. The dashed line is an illustration if the $100\%$ efficient case. RMS errors are indicated in the bottom right corners of each panel. For brighter stars (except for $14<g_mag<15$ see text) and for higher metallicity values, where the training set coverage is better, the estimation error is smaller.}
\label{fig:self}
\end{figure*}

\section{Evaluation of the photo-Met method} \label{sec:test}

\begin{figure}
\includegraphics[scale=0.92]{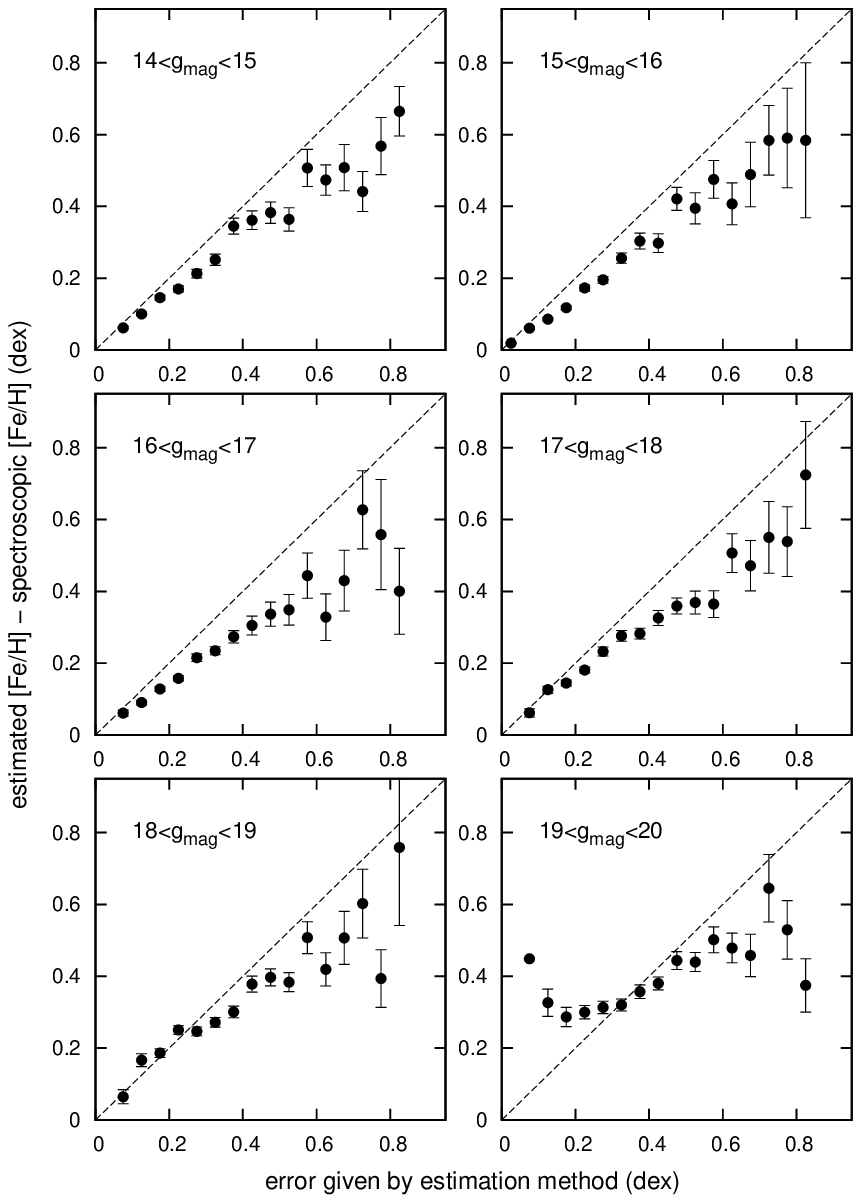}
\caption{Empirical errors vs. estimated errors, binned and averaged.}
\label{fig:error3}
\end{figure}

There are two major factors that affect the goodness of the estimation. As going towards fainter magnitudes, the photometric error increases and we expect our estimation to worsen as we apply it for fainter stars. To test this, we divided the training set into 6 equal sets based on the $g$-band magnitudes. After selecting 2000 stars randomly from each set for a test set and using the remaining stars as training set, we get 6 different ensembles. The results are shown in Fig.~\ref{fig:self}. RMS error is calculated in indicated in the panels. As we expect, the RMS increases as we go towards fainter samples, except for the first panel ($14<g_{mag}<15$). Images of stars with apparent magnitude of $g<14.5$ are often saturated \citep{Chonis2008}, thus have higher photometric uncertainties than the next two fainter sets.

It is also evident that the estimation works best in the high metallicity range. Towards lower metallicities, on the one hand, we get higher errors, on the other hand, the method tends to slightly overestimate the values. One reason is that, for this metallicity range, as Fig.~\ref{fig:TS} shows, the training set stars do not cover the parameter space densely enough. Another reason might be that the reference stars with small differences in metallicity are not separated well enough in the color space. We plan to investigate this latter effect in detail in a subsequent study. Altogether, there is a good chance that, for the new generation of surveys like Pan-STARRS and LSST, combining the higher photometric accuracy of the stacked observations with training sets covering a wide parameter range the performance of metallicity estimation can be improved.

Our technique yields an estimate not only on the metallicities but also on their errors. To analyse how reliable these estimations are, in Fig.~\ref{fig:error3} we plot the empirical errors in the form of 
\begin{equation*}
Q = \left| \left[ \text{Fe}/\text{H} \right]_\text{estimated} - \left[ \text{Fe} / \text{H} \right]_\text{spectroscopic} \right|
\end{equation*}
as a function of the estimated errors. For a better representation of the data, we binned the estimated errors into intervals of $0.05$~dex and took the average of the empirical errors. The ideal case, when the estimated errors exactly correspond to the empirical errors, is marked with a dotted line. The error bars indicate the standard deviation divided by the root of the number of points in each bin respectively. For most of the $g$ magnitude ranges the estimated error seems to be a good approximation of the true statistical error. We can see a somewhat similar trend as before, namely, by going to fainter $g$ magnitude ranges, the estimated errors tends to depart from the ideal, especially in the magnitude range of $g \simeq 19-20$.

\begin{figure}
\includegraphics[scale=0.92]{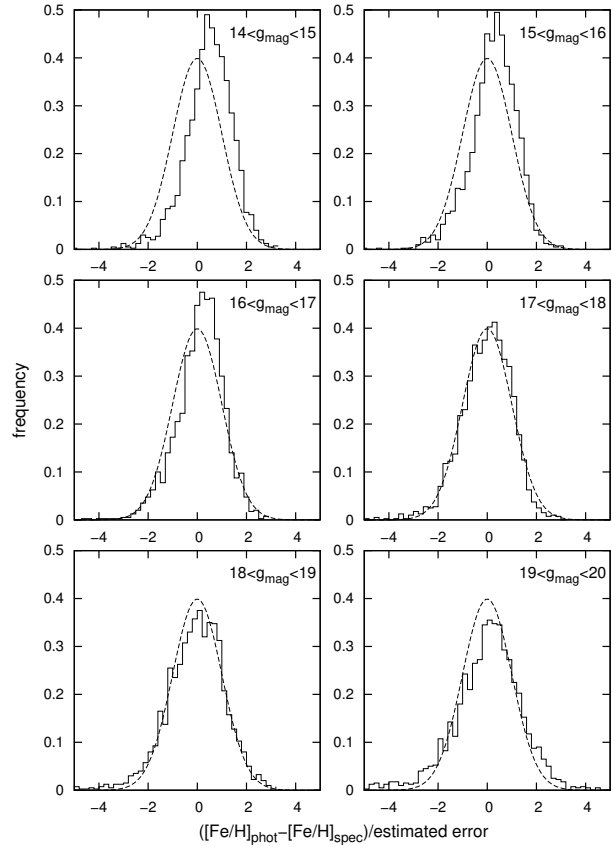}
\caption{The probability density distribution of normalized errors (solid line). As expected for unbiased estimators, the distributions are close to the normal distribution (dashed line).}
\label{fig:error_gauss}
\end{figure}

As we have already discussed in Sec.~\ref{sec:kd} the first step of our algorithm is when nearest neighbors are sought for every query point $p$. Let us consider the metallicity values ([Fe/H]$_i$) of the neighbours as realizations of identically distributed random variables. If we assume that they are independent then the probability distribution of the estimated metallicities if determined by the central limit theorem and, following the same chain of thought, also is the distribution of $Q$. If we divide $Q$ by the standard deviation of the error realizations (which, in this case, is the estimated error), the resulting quantity should follow a normal distribution, at least if the error estimates are correct \citep{Carliles2010}. This is just the case, in Fig.~\ref{fig:error_gauss} we show how the normalized empirical error fits to the standard normal distribution.

\section{Discussion} \label{sec:results}
The photo-Met method introduced in this study makes it possible to give fast and robust estimates on metallicity for a wide range of stars based solely on photometric observations and a spectroscopically confirmed training set. It performs better for stars with higher metal abundances and smaller photometric uncertainties. Beyond metal abundance estimation, it also yields reliable error estimates for every object. The uncertainty of metallicity estimates is reasonably good (especially in the $15 \leq g \leq 17$ range, where it is comparable to the spectroscopic measurements) and the probability distribution of the errors Gaussian. In the future, we will apply the technique for stars observed by the SDSS survey with no spectroscopic observations available. Based on the same photometric selection criteria as for the training set used in the present study, we estimate that reliable metallicities for about 19 million stars could be determined. By increasing the number of objects in the training set (especially in the region of lower abundances) and using better photometry from the upcoming surveys we can hope to further improve the quality of the technique in the future.

\acknowledgements %
The authors would like to acknowledge support from grants: NKTH: Pol\'anyi, KCKHA005 and OTKA-103244.

\end{document}